\documentclass[preprint,12pt]{elsarticle}
\usepackage{CJK,lineno}
\usepackage{amssymb}
\usepackage[pdftex,colorlinks]{hyperref}
\usepackage{multirow}
\usepackage[tbtags]{amsmath}

\biboptions{compress}

\begin{document}

\begin{frontmatter}

\title{A quantum algorithm to  approximate the linear structures of Boolean functions}

\author{Hong-Wei Li$^{1,2,3,4}$}
\author{Li Yang$^{1,3}$\corref{1}}
\cortext[1]{Corresponding author email: yangli@iie.ac.cn}
\address{1.State Key Laboratory of Information Security, Institute of Information Engineering, Chinese Academy of Sciences, Beijing 100093, China\\
2.School of Mathematics and Statistics, Henan Institute of Education, Zhengzhou,450046,Henan, China\\
3.Data Assurance and Communication Security Research Center, Chinese Academy of Sciences, Beijing 100093, China\\
4.University of Chinese Academy of Sciences, Beijing 100049, China}
\begin{abstract}
We present a quantum algorithm for approximating the linear structures of a Boolean function $f$. Different from previous algorithms (such as Simon's and Shor's algorithms) which rely on restrictions on the Boolean function, our algorithm applies to every Boolean function with no promise. Here, our methods are based on the result of the Bernstein-Vazirani algorithm which is to identify linear Boolean functions and the idea of Simon's period-finding algorithm. More precisely, how the extent of approximation changes over the time is obtained, and meanwhile we also get some quasi linear structures if there exists. Next, we obtain that the running time of the quantum algorithm to thoroughly determine this question is related to the relative differential uniformity $\delta_f$ of $f$. Roughly speaking, the smaller the $\delta_f$ is, the less time will be needed.
\end{abstract} 

\end{frontmatter}

\section{Introduction}	
\vspace*{-0.5pt}
\noindent

Linear structures of Boolean functions have important significance in cryptography \cite{LA94,XJ95,SD01,FX95}. Given a quantum oracle to a multiple output Boolean function $g$, under the promise that $g$  is one to one or $g$ has a nonzero period(i.e., a linear structure of $g$), Simon's algorithm\cite{DR97} could efficiently determine which is the case and find out the period if it has. Inspired by Simon's algorithm, Shor\cite{PW97} gave a polynomial-time algorithm for factoring integers. Both Simon's algorithm and Shor's algorithm have exponential speedups over the best known classical algorithms. However, \cite{RB01} pointed out that the exponential speedup could only be obtained for a problem with a promise in advance, any quantum algorithm for no restriction Boolean function could merely offer a polynomial speedup over the classical deterministic algorithm.

Recently, there were great interests in exploiting quantum algorithms to approximately solve some problems \cite{DVZ06,AR11,BHMT02,BDGT11}. In this paper, we mainly research the quantum algorithm for approximating the linear structures of a Boolean function $f$ with no promise at all.

The Deutsch-Jozsa algorithm \cite{DR92} and the Bernstein-Vazirani algorithm\cite{BV93} have the same network\cite{CEMM98}. Suppose $f$ has $n$ variables, if we run the same quantum network without the last measurement, the output will be a quantum state that is a superposition of all states $|w\rangle$ ($w\in \{0,1\}^n$), and the amplitude corresponding to each state $|w\rangle$ is its Walsh spectrum value $S_f(w)$. There have been some quantum algorithms for studying the properties of Boolean functions based on the Bernstein-Vazirani algorithm\cite{ME11,DEM13}. In addition, \cite{FP99} has shown a link between the Walsh spectrums and the linear structures of Boolean functions.

Inspired by the Bernstein-Vazirani algorithm and the conclusions in \cite{FP99}, we have an idea to do this work. First, we generate the results in \cite{FP99}, and then give our quantum algorithm, later apply the results we have got to analyse our algorithm.

\section{Preliminaries}
\subsection{The linearity of Boolean functions}
Let $n$ be a positive integer. $F_2=\{0,1\}$ denotes a finite field of characteristic 2, and $F_2^n$ is a vector space over $F_2$. A mapping from $F_2^n$ to $F_2$ is always called a Boolean function, and let $\mathfrak{B_n}$ denote the set of Boolean functions of $n$ variables. \\

\noindent{\bf Definition 1}\quad
A vector $a \in F^n_2$ is said to be a linear structure of a function $f\in \mathfrak{B_n}$ if
 \begin{equation}\label{eq:a}
  f(x\oplus a)+f(x)=f(a)+f(0), \,\,\forall x \in F^n_2,
  \end{equation}
 where $\oplus$ denotes bitwise exclusive-or, it is naturally the addition operation in $F^n_2$.

Let $U_f$ denote the set of the linear structures of $f$, and
\begin{equation}\label{eq:b}
U_f^{i}=\{a \in F^n_2|f(x\oplus a)+f(x)=i,\, \forall x \in F^n_2\}\,\, (i=0,1).
\end{equation}
Obviously $U_f=U_f^{0}\bigcup U_f^{1}.$

Let
\begin{equation}\label{eq:c}V_{f,a}^i=\{x \in F^n_2|f(x\oplus a)+f(x)=i\}\,(i=0,1),\,\,\forall a \in F^n_2.
\end{equation}

Let $|V|$ denote the cardinality of $V$. Obviously, $0\leqslant|V_{f,a}^i|/2^n\leqslant1$. $a\in U_f^{i}$ if and only if $|V_{f,a}^i|/2^n=1$. In this paper, we will use $1-|V_{f,a}^i|/2^n$ to describe the extent of a vector $a$ approximating to linear structure, and we hope it will be small enough. Naturally, we give the following two definitions.\\

\noindent{\bf Definition 2}\quad
A vector $a \in F^n_2$ is called a quasi linear structure of a function $f\in \mathfrak{B_n}$ if
\begin{equation}\label{eq:ab}
1-\frac{|\{x \in F^n_2|f(x\oplus a)+f(x)=i\}|}{2^n}<l(n),
\end{equation}
here the function $l(n)$ is negligible, more specifically, for any polynomial $p(\cdot)$, there exists an $N>0$ such that for all integers $n>N$, $l(n)<1/p(n)$ holds. That is to say, \eqref{eq:a} holds except a negligible number of $x$.\\

\noindent{\bf Definition 3}\quad The relative differential uniformity (usually say differential uniformity for abbreviation in this paper) of $f\in \mathfrak{B_n}$ is
\begin{equation}\label{eq:d}
\delta_f=\frac{1}{2^n}\max_{0\neq a\in F^n_2}\max_{i\in F_2}
|\{x \in F^n_2|f(x\oplus a)+f(x)=i\}|.
\end{equation}

Generally speaking, the $\delta_f$ given in \eqref{eq:d} satisfies $\frac{1}{2}\leq \delta_f\leq 1$.
$U_f\neq \{0\}$ if and only if $\delta_f=1$. \\

To study linear structures of a Boolean function, we define the  Walsh spectrum of it.\\

\noindent{\bf Definition 4}\quad Suppose $f\in \mathfrak{B_n}$, the Walsh spectrum of $f$ is defined as
\begin{equation}\label{eq:e}
S_f(w)=\frac{1}{2^n}\sum_{x\in F^n_2}(-1)^{f(x)+w\cdot x}.
\end{equation}

The following two theorems demonstrate the links between the Walsh spectrals and the linear structures, and we get inspirations from them.\\

\noindent{\bf Theorem A\cite{FP99}}\quad Suppose $f\in \mathfrak{B_n}$, the set of the linear structures of $f$ is $U_f=U_f^{0}\bigcup U_f^{1}$, then $a\in U_f^{i}(i=0,1)$ if and only if for $\forall w\in F^n_2: w\cdot a=\overline{i}=i+1$ (i.e. $w\cdot a\neq i$), $S_f(w)=0$. \\

\noindent{\bf Theorem B\cite{FP99}}\quad
\begin{equation}\label{eq:ea}
U_f^{0}=\{\alpha\in F^n_2|\beta\cdot\alpha=0,
\forall \beta\in\{w\in F^n_2|w=\sum_jw_j,\,S_f(w_j)\neq 0\}\}.
\end{equation}

We will generate these two theorems in the following section, and ours contain them.
For convenience, let
\begin{equation}\label{eq:f}
N_f^0=\{w\in F^n_2|S_f(w)=0\},\qquad N_f^1=\{w\in F^n_2|S_f(w)\neq 0\}.
\end{equation}

\subsection{The Bernstein-Vazirani algorithm \cite{BV93,CEMM98} }

The Bernstein-Vazirani algorithm is to distinguish linear functions. Specifically, suppose
\begin{equation}\label{eq:g}
f(x)=a\cdot x=\sum_{i=1}^n a_ix_i.
\end{equation}
The algorithm aims to determine $a$. We give a description about how the algorithm works as follows.

1. Perform the Hadamard transform $H^{(n+1)}$ on the initial state $|\psi_0\rangle=|0\rangle^{\otimes n}|1\rangle$, giving
\begin{equation}\label{eq:h}|\psi_1\rangle=\sum_{x\in F^n_2}
\frac{|x\rangle}{\sqrt{2^n}}\cdot\frac{|0\rangle-|1\rangle}{\sqrt{2}}.\end{equation}

2. Apply the $f$-controlled-NOT gate on $|\psi_1\rangle$, producing
\begin{equation}\label{eq:i}
|\psi_2\rangle=\sum_{x\in F^n_2}
\frac{(-1)^{f(x)}|x\rangle}{\sqrt{2^n}}\cdot\frac{|0\rangle-|1\rangle}{\sqrt{2}}.
\end{equation}

3. We again apply $n$ Hadamard gates to the first $n$ qubits yielding
\begin{equation}\label{eq:j}
|\psi_3\rangle =\sum_{y\in F^n_2}\frac{1}{2^n}\sum_{x\in F^n_2}(-1)^{f(x)+y\cdot x}|y\rangle
\cdot\frac{|0\rangle-|1\rangle}{\sqrt{2}}.
\end{equation}
If $f(x)=a\cdot x$,
\begin{equation}\label{eq:ac}
\begin{split}
|\psi_3\rangle
&=\sum_{y\in F^n_2}\chi_a(y)|y\rangle
\cdot\frac{|0\rangle-|1\rangle}{\sqrt{2}}\\
&=|a\rangle\cdot\frac{|0\rangle-|1\rangle}{\sqrt{2}},
\end{split}
\end{equation}
where
\begin{equation}\label{eq:k}
\chi_a(y)=
\begin{cases}
1& \text{if $y=a$},\\
0& \text{if $y\neq a$}.
\end{cases}
\end{equation}
Now we measure the first $n$ qubits of $|\psi_3\rangle$ in the computational basis, we find $a$ with probability 1.

If $f(x)\in \mathfrak{B_n}$ is not linear, run the Bernstein-Vazirani algorithm, the output can be expressed as
\begin{equation}\label{eq:l}
|\psi_3\rangle=\sum_{y\in F^n_2}S_f(y)|y\rangle
\cdot\frac{|0\rangle-|1\rangle}{\sqrt{2}}
\end{equation}
according to  \eqref{eq:e} and \eqref{eq:j}. And then we measure the first $n$ qubits in the computational basis, we find $y$ with probability $(S_f(y))^2$ (we will write it as $S_f^2(y)$ for convenience). That is, if we repeat the algorithm time and again, we will obtain the $y\in N_f^1$ all the time. It will be helpful when we take account of the linear structures of Boolean functions.

\section{The further relationships between the linear structures and Walsh spectrums of Boolean functions}
The following theorems play a pivotal role in applying the quantum algorithm to seek the linear structures of Boolean functions. They build a bridge between the linear structures and the Walsh spectrums of Boolean functions so that we can use the Bernstein-Vazirani algorithm to solve the problem. Compare with the previous two theorems, ours list here are more specific. Theorem A is qualitative, the following one is quantitative.

\vspace*{12pt}
\noindent
{\bf Theorem 1}\quad
Let $f\in \mathfrak{B_n}$, then $\forall a\in F^n_2$, $\forall i\in F_2$,
\begin{equation}\label{eq:m}
\sum_{w\cdot a=i}S_f^2(w)=\frac{|V_{f,a}^i|}{2^n}
=\frac{|\{x \in F^n_2|f(x\oplus a)+f(x)=i\}|}{2^n}.
\end{equation}

Theorem 1 demonstrates if we run the Bernstein-Vazirani algorithm,  the probability of getting $w$ with $w\cdot a=i$ will be equal to $\frac{|V_{f,a}^i|}{2^n}$ . To prove Theorem 1, we will need the following lemma appearing in \cite{FP99}, and we will give a proof of it in appendix.

\vspace*{12pt}
\noindent
{\bf Lemma 1}\quad
\begin{equation}\label{eq:ab}
C_f(a)=\sum_{x\in F^n_2}(-1)^{f(x)+f(x\oplus a)}=2^n(\sum_{w\cdot a=0}S^2_f(w)-\sum_{w\cdot a=1}S^2_f(w)),
\end{equation}
where $C_f(a)$ is the correlation function of $f$, and $-$ is the subtraction operation of the integer ring.

\vspace*{12pt}
\noindent
{\bf Proof of Theorem 1}  \quad First of all,
\begin{equation}\label{eq:p}
\begin{split}
C_f(a)&=|\{x \in F^n_2|f(x\oplus a)+f(x)=0\}|-
|\{x \in F^n_2|f(x\oplus a)+f(x)=1\}|\\
&=|V_{f,a}^0|-|V_{f,a}^1|,
\end{split}
\end{equation}
From \eqref{eq:ab} and \eqref{eq:p}, we have
\begin{equation}\label{eq:q}
\sum_{w\cdot a=0}S^2_f(w)-\sum_{w\cdot a=1}S^2_f(w)=
\frac{|V_{f,a}^0|}{2^n}-\frac{|V_{f,a}^1|}{2^n}.
\end{equation}
In addition, Parseval's relation gives
\begin{equation}\label{eq:r}
\sum_{w\cdot a=0}S^2_f(w)+\sum_{w\cdot a=1}S^2_f(w)=
\sum_{w\in F^n_2}S^2_f(w)=1.
\end{equation}
And by the definition of $V_{f,a}^i$, we have
\begin{equation}\label{eq:s}
|V_{f,a}^0|+|V_{f,a}^1|=2^n.
\end{equation}
From \eqref{eq:r} and \eqref{eq:s}, we obtain
\begin{equation}\label{eq:t}
\sum_{w\cdot a=0}S^2_f(w)+\sum_{w\cdot a=1}S^2_f(w)=
\frac{|V_{f,a}^0|}{2^n}+\frac{|V_{f,a}^1|}{2^n}.
\end{equation}
Combining \eqref{eq:q} and \eqref{eq:t}, we achieve
\begin{equation}\label{eq:u}
\begin{cases}
\sum_{w\cdot a=0}S^2_f(w)=\frac{|V_{f,a}^0|}{2^n},\\
\sum_{w\cdot a=1}S^2_f(w)=\frac{|V_{f,a}^1|}{2^n}.
\end{cases}
\end{equation}
\eqref{eq:u} is essentially the same as \eqref{eq:m}.

As an application, we have the following theorem.

\vspace*{12pt}
\noindent
{\bf Theorem 2}\quad
Let $f\in \mathfrak{B_n}$, then $\forall i\in \{0,\,1\}$,
\begin{equation}\label{eq:v}
U_f^i=\{a\in F^n_2|w\cdot a=i,\,\forall w\in N_f^1\}.
\end{equation}

\vspace*{12pt}
\noindent
{\bf Proof of Theorem 2} \quad   Recall that if and only if is usually abbreviated to iff in mathematics.
By the definition of $U_f^i$ and $V_{f,a}^i$, we have $a\in U_f^i$ iff $|V_{f,a}^i|=2^n$ and $|V_{f,a}^{\overline{i}}|=0.$
Reference to \eqref{eq:u}, this holds iff $\sum_{w\cdot a=i}S^2_f(w)=1$ and $\sum_{w\cdot a=\overline{i}}S^2_f(w)=0$.
In other words, $\forall w\in F^n_2,\: \text{as long as}\: w\cdot a=\overline{i}, \:\text{it will be} \: S_f(w)=0;$ and $\forall w\in F^n_2,\: \text{as long as}\: S_f(w)\neq 0, \:\text{it will be}\: w\cdot a=i.$
This is in fact equivalent to
\begin{equation*}
a\in \{a\in F^n_2|w\cdot a=i,\,\forall w\in N_f^1\}.
\end{equation*}

From Theorem 2, if we can get the set $N_f^1$, we will obtain $U_f^i$. Moreover, we have known that repeating the Bernstein-Vazirani algorithm will give a subset of $N_f^1$.

\section{The quantum algorithm for the linear structures of Boolean Functions}

We will now state a quantum algorithm to decide whether a function has non-zero linear structures or not. If the differential uniformity $\delta_f$ of $f\in \mathfrak{B_n}$ is no more than a constant $\delta$ ($\frac{1}{2}\leqslant\delta<1$ is independent to $n$), it will definitely give "no." If $1-\frac{1}{e(n)}\leqslant\delta_f\leqslant 1$ ($e(n)$ is a exponential function of $n$), it will give "yes" with a great probability and give quasi linear structures. These quasi linear structures may be the real ones, and also may be the approximate ones.

\subsection{ The quantum algorithm}

Our algorithm is based on the Bernstein-Vazirani algorithm. Furthermore, we solve a system of linear equations as Simon's algorithm does. The details of the algorithm are presented as following.

\begin{bframe}
\textbf{Algorithm 1}

Suppose $p(n)$ is an arbitrary polynomial function of $n$, $\Phi$ is null.

1.\;Initialize $H:=\Phi$, $r:= p(n)$.

2.\;For $r$ many times do

\quad 2.1.\;Run the Bernstein-Vazirani algorithm to the function $f$ for $n+1$ times to get $n+1$ vectors $w_1, \cdots, w_{n+1} \in N_f^1$.

\quad 2.2.\;Update $H:= H\bigcup\{w_1, \cdots, w_{n+1}\}$.

\quad 2.3.\;Solve the equations $x\cdot H=i$ to get the solution $A^i$.

\quad 2.4.\;If $A^0=\{0\}$ and $A^1=\Phi$, then output no and halt.

3.\;Report $f$ has quasi linear structures and output $A^0$ and $A^1$.

\end{bframe}

It must be $U_f^i\subseteq A^i$, but it may not be $U_f^i\supseteq A^i$. So if $A^0=\{0\}$ and $A^1=\Phi$($\Phi$ is a null set), the function $f$ is declared to have no non-zero linear structure.

Particularly, we have the following helpful results which will be proved in appendix.
If $0\in H$, or $\exists$ even numbers of $w_j$ such that $\sum_jw_j\in H$, then $A^1=\Phi$, thereby $U_f^1=\Phi$. If there are $k=n$ linearly independent vectors in $H$, then $A^0=\{0\}$, $|A^1|=1$ or $|A^1|=0$, accordingly $U_f^0=\{0\}$.

\subsection{ The analysis of the above quantum algorithm}

Now we think about the questions below:
How many times should the algorithm be repeated to get the conclusion?
If we run the algorithm at most polynomial times, what we get? Naturally,
we can get the conclusion that a function $f\in \mathfrak{B_n}$ has no non-zero linear structure or else we can't. If we can't, that is, we find out some approximate or exact linear structures through the algorithm, what properties should these vectors possess?
What is the running time of the quantum algorithm to thoroughly determine this question?
The following are the answers to these questions.

\vspace*{12pt}
\noindent
{\bf Theorem 3 }\quad Given an oracle access to a Boolean function $f$ with $n$ variables, Algorithm 1 gives an answer that $f$ has no non-zero linear structure or outputs  vector sets $A^0$ and $A^1$. Notice that Algorithm 1 actually repeats the Bernstein-Vazirani algorithm $m=r\cdot(n+1)$ times, we have $\forall a\in A^i(i=0,1)$, $\forall \epsilon$, $0<\epsilon<1$,
\begin{equation}
\text{Pr}(1-\frac{|\{x \in F^n_2|f(x\oplus a)+f(x)=i\}|}
{2^n}<\epsilon)>1-e^{-2m\epsilon^2},
\end{equation}
here $\text{Pr}(F)$ denotes the probability of the event $F$ happens.

It demonstrates that if $l(n)>0$ is a polynomial function of $n$, $\epsilon=\frac{1}{l(n)}$, and $r=p(n)=l^2(n)$,
the probability will be very close to 1. Thus those vectors in set $A=A^0\cup A^1$ are quasi linear structures except a negligible probability. In other words, $\forall a\in A^i (i=0,1)$, $(a,i)$ is a high probability differential.

\vspace*{12pt}
\noindent
{\bf Proof of Theorem 3} \quad  For any $a^i\in A^i(i=0,1)$,
\begin{equation}
\text{Pr}(f(x\oplus a^i)+f(x)=i)
=\frac{|\{x\in F_2^n|f(x\oplus a^i)+f(x)=i\}|}{2^n}
=\frac{|V_{f,a^i}^i|}{2^n}.
\end{equation}
Let
\begin{equation}
\frac{|V_{f,a^i}^i|}{2^n}=p,\qquad\frac{|V_{f,a^i}^{\overline{i}}|}{2^n}=1-p=q.
\end{equation}
Obviously  $p,q\in[0,1]$. And let $Y$ be a random variable
\begin{equation}
Y(w)=
\begin{cases}
0 & w\cdot a^i=i\\
1 & w\cdot a^i=\overline{i}
\end{cases}
\end{equation}
then from Theorem 1, the expectation of $Y$ is $\text{E}(Y)=1\cdot q=q=1-p$, $m$ times running the Bernstein-Vazirani algorithm correspond to $m$ independent identical random variables $Y_1,\cdots, Y_m$.
By Hoeffding's inequality\cite{WH63},
\begin{equation}\label{eq:y}
\text{Pr}(q-\frac{1}{m}\sum_{j=1}^mY_j\geqslant\epsilon)\leqslant e^{-2m\epsilon^2}.
\end{equation}
Now that one has got $a^i$, $\sum_{j=1}^mY_j$ in \eqref{eq:y} must equal $0$ (because if there exists some $Y_j=1$, we can't get $a^i\in A^i$). Hence
\begin{equation}\label{eq:z}
\text{Pr}(q\geqslant\epsilon)\leqslant e^{-2m\epsilon^2}.
\end{equation}
From \eqref{eq:z} and $q=1-p$, also $p$ is always no more than 1, we have
\begin{equation}
\text{Pr}(1-p<\epsilon)=\text{Pr}(1-\epsilon<p\leqslant 1)>1-e^{-2m\epsilon^2}.
\end{equation}
That is to say, $(1-\frac{1}{m^\lambda},1](0<\lambda\leqslant\frac{1}{2})$ is the confidence interval of $p$ with confidence level $1-e^{-2m^{1-2\lambda}}$(here let $\epsilon=\frac{1}{m^\lambda}$).

\vspace*{12pt}
\noindent
{\bf Theorem 4}\quad To thoroughly determine whether a given Boolean function $f$ has non-zero linear structures or not, the $r$ in Algorithm 1 should rely on the differential uniformity $\delta_f$ of $f$. More precisely, if $\delta_f<1$, an average of $O(\frac{n+1}{1-\delta_f})$ times running the Bernstein-Vazirani algorithm (i.e. $r=O(\frac{1}{1-\delta})$ in Algorithm 1) will give an answer that $f$ has no non-zero linear structure.

\vspace*{12pt}
\noindent
{\bf Proof of Theorem 4} \quad If $\exists \xi\in F_2^n$, $\mu\in F_2$, such that
\begin{equation}
\begin{split}
\delta_f&=\frac{1}{2^n}
|\{x \in F^n_2|f(x\oplus \xi)+f(x)=\mu\}|\\
&=\frac{1}{2^n}\max_{0\neq a\in F^n_2}\max_{i\in F_2}
|\{x \in F^n_2|f(x\oplus a)+f(x)=i\}|\\
&=\delta<1,
\end{split}
\end{equation}
then
\begin{equation}
\begin{split}
&\quad\,\frac{1}{2^n}
|\{x \in F^n_2|f(x\oplus \xi)+f(x)=\overline{\mu}\}|\\
&=\frac{1}{2^n}\min_{0\neq a\in F^n_2}\min_{i\in F_2}
|\{x \in F^n_2|f(x\oplus a)+f(x)=i\}|\\
&=1-\delta>0.
\end{split}
\end{equation}
Therefore $\forall a\in F^n_2$, $a\neq 0$, $\forall i\in F_2$,
\begin{equation}\label{eq:ac}
0<1-\delta\leqslant\frac{1}{2^n}|\{x \in F^n_2|f(x\oplus a)+f(x)=i\}|\leqslant \delta<1,
\end{equation}
let $B_{a,i}=\{w\in F^n_2|w\cdot a=i\}$.
Suppose one has repeated the Bernstein-Vazirani algorithm for $m$ times, and has obtained $H$. By theorem 1, the probability that $H\subseteq B_{a,i}$ is at most
\begin{equation}
\delta^m=(1-(1-\delta))^m\leqslant e^{-m(1-\delta)}.
\end{equation}
That can be made small if we choose $m$ of order $(1-\delta)^{-1}$ (i.e. $m=O(\frac{1}{1-\delta})$). In particular, if
\begin{equation}
m>\frac{c}{1-\delta},
\end{equation}
here $c>1$ is a constant, then
\begin{equation}
\delta^m<\frac{1}{e^c}.
\end{equation}
Therefore the probability that $H\nsubseteq B_{a,i}$ is at least
\begin{equation}
1-\delta^m>1-\frac{1}{e^c}.
\end{equation}

Learning from \cite{GP97}, suppose we have got $k(0\leqslant k<n)$ linearly independent vectors $w_1,\cdots,w_k$ at some time, and $a\in F_2^n$ is a solution of the equations $w_j\cdot x=0(j=1,\cdots,k)$, then after another expected number of order $O(\frac{1}{1-\delta})$ measures, we will get a $w_{k+1}$ with $w_{k+1}\cdot a=1$. This $w_{k+1}$ must be linearly independent with $w_1,\cdots,w_k$, since it should be $w\cdot a=0$ for any linear combination $w$ of $w_1,\cdots,w_k(j=1,\cdots,k)$. As a result, we would find out $n$ linearly independent vectors $w_1,w_2\cdots,w_n$ through an expected number of order $O(\frac{n}{1-\delta})$ measures. From this fact, we can know $U_f^0=\{0\}$. About the only one possible solution $b$ of the equations $w_j\cdot x=1$, $j\in\{1,2,\cdots,n\}$, another expected number of order $O(\frac{1}{1-\delta})$ measures will give a $w_{n+1}$ with $w_{n+1}\cdot b=0$, this shows $U_f^1=\Phi$. Consequently, through $m=O(\frac{n+1}{1-\delta_f})$ times running we get $U_f=\{0\}$.

\vspace*{12pt}

Generally, if $\delta<1-\frac{1}{p(n)}$($p(n)$ is a polynomial function of $n$), by Theorem 4, through $O(\frac{n+1}{1-\delta})<O((n+1)p(n))$ times running the Bernstein-Vazirani algorithm (i.e. $r=O(p(n))$ in Algorithm 1) will give a "no" answer. A special case is the Bent function, whose differential uniformity is $\delta=\frac{1}{2}$, through $O(2n)$ times will do. If $1>\delta>1-\frac{1}{e(n)}$($e(n)$ is an exponential function of $n$), by Theorem 3 and Theorem 4, at least $O(e(n))$ times running will be needed to give the exact result. A special case is $\delta=1-\frac{1}{2^{n-1}}$, it need $O(n2^{n-1})$ times. If after $O(n2^{n})$ times running the algorithm, it still has quasi linear structures, these ones must be linear structures except a negligible probability. Any polynomial time would give an approximate solution which is actually the high probability differential.

\section{Conclusions}
While the best known classical algorithm for computing Walsh spectral is of order $O(n2^n)$, the quantum algorithm (specifically the Bernstein-Vazirani algorithm) can give some informations of it efficiently. Based on this,
We present a polynomial-time quantum approximate algorithm for the linear structure of the Boolean functions. The quantum algorithm can give a "no" answer or an approximate solution set (i.e. quasi linear structures set) which includes and tends to the linear structures set. We haven't seen any classical efficient algorithm to this question. Besides, We have also analyzed the efficiency of the algorithm. The error range with the error probability is given. To thoroughly solve the question, the running time of the algorithm is linked to the differential uniformity of the function. The smaller of the differential uniformity, the less time should be needed. Just like some papers \cite{ME11,DEM13}, we can use the Grover like operator to amplify the amplitude to get a better conclusion, but that is not our concerns. The point is that maybe the quasi linear structures we get can guide the differential cryptanalysis. And also the results elaborated in this paper verify the conclusion in \cite{RB01}.

\section*{Acknowledgement}

 This work was supported by the National Natural Science Foundation of China under Grant No.61173157.


\vspace*{12pt}

For the self-contained of this paper, we give the following appendixes.

\vspace*{12pt}
\noindent{\textbf{Appendix}}

\vspace*{12pt}
\noindent
{\bf Lemma 1}\quad
\begin{equation*}
C_f(a)=\sum_{x\in F^n_2}(-1)^{f(x)+f(x\oplus a)}=2^n(\sum_{w\cdot a=0}S^2_f(w)-\sum_{w\cdot a=1}S^2_f(w)),
\end{equation*}
\noindent{\bf  Proof of the Lemma 1}\qquad
\begin{equation*}
\begin{split}
C_f(a)&=\sum_{x\in F^n_2}(-1)^{f(x)+f(x\oplus a)}\\
&=\sum_{x\in F^n_2}(\sum_{w\in F^n_2}S_f(w)(-1)^{w\cdot x})
\cdot(\sum_{w\in F^n_2}S_f(w)(-1)^{w\cdot (x\oplus a)})\\
&=\sum_{x\in F^n_2}(\sum_{w\cdot a=0}S_f(w)(-1)^{w\cdot x}+
\sum_{w\cdot a=1}S_f(w)(-1)^{w\cdot x})\cdot\\
&\qquad\quad(\sum_{w\cdot a=0}S_f(w)(-1)^{w\cdot x}-
\sum_{w\cdot a=1}S_f(w)(-1)^{w\cdot x})\\
&=\sum_{x\in F^n_2}[(\sum_{w\cdot a=0}S_f(w)(-1)^{w\cdot x})^2-
(\sum_{w\cdot a=1}S_f(w)(-1)^{w\cdot x})^2]\\
&=\sum_{x\in F^n_2}\sum_{w\cdot a=0}\sum_{\eta\cdot a=0}
S_f(w)S_f(\eta)(-1)^{(w\oplus \eta)\cdot x}\\
&\quad-\sum_{x\in F^n_2}\sum_{w\cdot a=1}\sum_{\eta\cdot a=1}
S_f(w)S_f(\eta)(-1)^{(w\oplus \eta)\cdot x}\\
&=\sum_{w\cdot a=0}\sum_{\eta\cdot a=0}
S_f(w)S_f(\eta)2^n\chi_w(\eta)\\
&\quad-\sum_{w\cdot a=1}\sum_{\eta\cdot a=1}
S_f(w)S_f(\eta)2^n\chi_w(\eta)\\
&=2^n(\sum_{w\cdot a=0}S^2_f(w)-\sum_{w\cdot a=1}S^2_f(w)).
\end{split}
\end{equation*}

From Theorem 2, according to the knowledge about the solutions of the linear equations in algebra, we have the following conclusions.

\vspace*{12pt}
\noindent
{\bf Proposition 1\footnote{Proposition 1 also appears in \cite{FP99}.}}\quad
If $0\in N_f^1$, i.e. $S_f(0)\neq 0$, then $U_f^1=\Phi$.

\vspace*{12pt}
\noindent
{\bf Proof} \quad If $0\in N_f^1$, the equation $0\cdot x=1$ has no solution, so $U_f^1=\Phi$.

\vspace*{12pt}
\noindent
{\bf Proposition 2}\quad
If $\exists w_1, w_2\in N_f^1$, and $w_1+w_2\in N_f^1$, then $U_f^1=\Phi$.

\vspace*{12pt}
\noindent
{\bf Proof} \quad If $w_1, w_2, w_1+w_2\in N_f^1$, the equation set
\begin{equation*}
\begin{cases}
w_1\cdot x=1\\
w_2\cdot x=1\\
(w_1\oplus w_2)\cdot x=1
\end{cases}
\end{equation*}
has no solution, so $U_f^1=\Phi$.

\vspace*{12pt}
\noindent
{\bf Proposition 3}\quad
If $dim N_f^1=k(k\leqslant n)$, then $dim U_f^0=n-k$. And if $U_f^1\neq \Phi$,
then $|U_f^1|=|U_f^0|$, and $dim U_f=n-k+1$.

\vspace*{12pt}
\noindent
{\bf Proof} \quad If $dim N_f^1=k(k\leqslant n)$, suppose $|N_f^1|=N$, and
$N_f^1=\{w_1, w_2, \cdots, w_N\}$. Then the equation set
\begin{equation}\label{eq:w}
\begin{cases}
w_1\cdot x=0\\
w_2\cdot x=0\\
w_N\cdot x=0
\end{cases}
\end{equation}
has the solution space of dimensions $n-k$. Suppose $A_f^0=\{a_1^0, \cdots, a_{2^{n-k}}^0\}$ is the solution set of the equations \eqref{eq:w}, then by Theorem 2, $U_f^0=A_f^0$. Suppose $b$ is a special solution of the equation system
\begin{equation}\label{eq:x}
\begin{cases}
w_1\cdot x=1\\
w_2\cdot x=1\\
w_N\cdot x=1
\end{cases}
\end{equation}
Then $A_f^1=\{a_1^0\oplus b, \cdots, a_{2^{n-k}}^0\oplus b\}$ is the solution set of the equations \eqref{eq:x}. By Theorem 2, $U_f^1=A_f^1$. Therefore, $|U_f^1|=|U_f^0|$, and $dim U_f=n-k+1$.

\vspace*{12pt}
\noindent
{\bf Proposition 4}\quad
Even if $dim N_f^1=n$, there might be $U_f^1\neq \Phi$.

\vspace*{12pt}
\noindent
{\bf Proof} \quad For example,
\begin{equation}\label{eq:ax}
f(x_1x_2x_3)=x_1+x_2+x_1x_2+x_2x_3+x_1x_3,
\end{equation}
\begin{equation*}
S_f(001)=-\frac{1}{2}, S_f(010)=S_f(100)=S_f(111)=\frac{1}{2};
\end{equation*}
\begin{equation}\label{eq:bx}
S_f(000)=S_f(011)=S_f(101)=S_f(110)=0.
\end{equation}
\begin{equation}\label{eq:cx}
f(x_1x_2x_3\oplus111)=x_1+x_2+x_1x_2+x_2x_3+x_1x_3+1=f(x_1x_2x_3)+1.
\end{equation}

\end{document}